\documentstyle[aps,prl,preprint]{revtex}

\newcommand{\diag}{\mbox{\rm diag}}
\newcommand{\real}{\mbox{\rm Re}}

\newcommand{\trace}{\mbox{\rm Tr}}
\newcommand{\one}{\openone}
\def\nl{\nonumber \\}

\tightenlines

\begin{document}
\preprint{\sc FISIST/19-99/CFIF}
\draft


\title{Texture Zeros and Weak Basis Transformations}
\author{G.C. Branco\footnote{Email address: d2003@beta.ist.utl.pt},
 D. Emmanuel-Costa\footnote{Email address: david@gtae2.ist.utl.pt} and
 R. Gonz\'{a}lez Felipe\footnote{Email address: gonzalez@gtae3.ist.utl.pt}}
\address{Centro de F\'{\i}sica das Interac\c{c}\~{o}es Fundamentais, \\
Departamento de F\'{\i}sica, Instituto Superior T\'{e}cnico}
\address{Av. Rovisco Pais, 1096 Lisboa Codex, Portugal}
\date{November 19, 1999}
\maketitle

\begin{abstract}
We investigate the physical meaning of some of the ``texture zeros" which
appear in most of the \emph{Ans\"{a}tze} on quark masses and mixings. It is shown
that starting from arbitrary quark mass matrices and making a suitable weak
basis transformation one can obtain some of these sets of zeros which
therefore have no physical content. We then analyse the physical implications
of a four-texture zero \emph{Ansatz} which is in agreement with all present
experimental data.
\end{abstract}



\section{Introduction}

Understanding the structure of quark masses and mixings is one of the major
open questions in particle physics. Recently, one has had increasingly more
accurate experimental data \cite{pdg} on the value of quark masses, as well as
on the Cabibbo-Kobayashi-Maskawa (CKM) matrix \cite{ckm}. From this wealth of
experimental data, one may be tempted to extract a clue about a symmetry
principle, probably imposed at high-energy scale, which could lead to the
pattern of quark masses and mixings, observed at low-energy scale. One of the
difficulties one encounters in implementing this bottom-up approach results
from the large redundancy contained in the quark mass matrices within the
framework of the Standard Model (SM). In the SM, the flavour structure of
Yukawa couplings is not constrained by gauge symmetry and, as a result, the up
and down quark mass matrices are arbitrary complex matrices, thus containing a
total of 36 free parameters. This number is to be compared to the ten physical
parameters corresponding to the six quark masses and four physical parameters
of the CKM matrix. The above redundancy is closely related to the fact that
one has the freedom to make weak-basis (WB) transformations under which the
quark mass matrices change but the gauge currents remain diagonal and real.
Two sets of quark mass matrices related by a WB transformation obviously have
the same physical content.

In the literature, various approaches to the flavour puzzle have been
pursued, including the systematic search for texture zeros \cite{rrr}.
In view of the freedom in the choice of WB, it is important to analyse
within the SM when does a set of zeros results only from a choice of
WB and when it does imply restrictions among quark masses and/or
mixings.

In this paper, we will analyse systematically the physical content of
texture zeros within the framework of hermitian quark mass
matrices. At this point, it is worth recalling that, within the SM,
starting from arbitrary quark mass matrices $\tilde{M_u}$,
$\tilde{M_d}$, one can always make a WB transformation under which
$\tilde{M_u}\rightarrow M_u$, $\tilde{M_d}\rightarrow M_d$, with
$M_u$, $M_d$ hermitian matrices. Therefore, one does not loose
generality by restricting the analysis to hermitian quark mass
matrices.

We will consider some of the texture zeros which appear in most of the
\emph{Ans\"{a}tze} proposed in the literature
\cite{rrr,weinberg,wilczek,fritzsch,gatto,albright,georgi,dimopoulos},
namely those occurring in the elements (1,1), (1,3), (3,1) of the up
and down quark mass matrices. We will show which combinations of
texture zeros can be obtained from arbitrary quark mass matrices by
simply making appropriate WB transformations. Zeros satisfying the
above condition will be called WB zeros.  Of course, these sets of WB
zeros do not have any physical content by themselves. However, we will
point out that some of these sets, when supplemented by some mild
assumptions on the hierarchy of quark masses and mixings, do lead to
physical predictions.

\section{Zeros arising from WB transformations}

We will consider WB zeros occurring in the elements (1,1), (1,3) of
the up and down quark mass matrices. Since we restrict ourselves to
hermitian mass matrices , the vanishing of the (1,3) element obviously
implies the vanishing of the (3,1) element. Taking into account that
there are ten physical parameters in $M_u$, $M_d$, it is clear that
there is a limit to the number of WB zeros which can be obtained,
since the resulting $M_u$, $M_d$ matrices have to contain at least ten
independent parameters. Matrices with less than ten parameters lead to
relations between quark masses and/or mixings and therefore cannot
reflect only a choice of WB. In view of this, we need to consider
hermitian matrices containing at most four WB zeros, corresponding to
ten physical parameters. We will first consider matrices with WB zeros
in the (1,1) element of the up and down quark mass matrices and then
study the simultaneous presence of WB zeros in the (1,1) and (1,3)
elements.

\subsection{The (1,1) WB zero}

The most general WB transformation that leaves the mass matrices
hermitian is:

\begin{eqnarray}
M_u &\longrightarrow M'_u=W^{\dagger}\,M_u\,W,\nonumber\\ M_d
&\longrightarrow M'_d=W^{\dagger}\,M_d\,W, \label{transf1}
\end{eqnarray}
where $W$ is an arbitrary unitary matrix. In such basis, the quark
mass matrices can be diagonalized by the set of unitary matrices
$\{U_u,U_d\}$ such that

\begin{eqnarray}
D_u=U^{\dagger}_u\,M_u\,U_u,\nonumber\\ D_d=U^{\dagger}_d\,M_d\,U_d,
\label{diag}
\end{eqnarray}
where $D_u \equiv \diag(m_u,m_c,m_t)$ and $D_d \equiv
\diag(m_d,m_s,m_b)$. We emphasize that the relative sign of the quark
mass parameters $m_i$ $(i=u,c,t,d,s,b)$ does not have physical meaning
since it can always be changed by a WB transformation.

Let us start by choosing a WB where the up quark mass matrix $M_u$ is
diagonal and the down quark mass matrix $M_d$ is hermitian, i.e.

\begin{eqnarray}
M_u&=&D_u, \nonumber \\%
M_d&=&V\,D_d\,V^{\dagger}. \label{mudiag}
\end{eqnarray}
The matrix $V$ is an arbitrary unitary matrix, which in a physical context
would correspond to the usual CKM matrix.

Since we are interested in a WB where the (1,1)-matrix elements of the
quark mass matrices vanish, let us make a WB transformation W under
which $M_u$, $M_d$ transform as:

\begin{eqnarray}
M_u &\longrightarrow & M'_u=W^{\dagger}\,D_u\,W, \nonumber\\
M_d&\longrightarrow& M'_d=W^{\dagger}\,V\,D_d\,V^{\dagger}\,W, \label{WBT1}
\end{eqnarray}
such that $(M'_u)_{11}=(M'_d)_{11}=0.$ This requires the solution of the
system of equations
\begin{eqnarray}
m_u\:|W_{11}|^2+m_c\:|W_{21}|^2+m_t\:|W_{31}|^2=0,\nonumber\\
m_d\:|X_{11}|^2+m_s\:|X_{21}|^2+m_b\:|X_{31}|^2=0,\nonumber\\
|W_{11}|^2+|W_{21}|^2+|W_{31}|^2=1,\label{system1}
\end{eqnarray}
where $X \equiv V^{\dagger}W$, and thus:

\begin{eqnarray}
|X_{i1}|^2 & = & |V_{1i}|^2 |W_{11}|^2+|V_{2i}|^2
 |W_{21}|^2+|V_{3i}|^2 W_{31}|^2 \nl \nl & & + 2 \: \real(V^{\ast}_{1i}
 W_{11} V_{2i} W^{\ast}_{21}) + 2 \: \real(V^{\ast}_{1i} W_{11} V_{3i}
 W^{\ast}_{31}) \nl \nl & & + 2 \: \real(V^{\ast}_{2i} W_{21} V_{3i}
 W^{\ast}_{31}), \quad (i=1,2,3).
\label{X}
\end{eqnarray}

Notice that for the system (\ref{system1}) to have a real solution, it
is necessary that at least one of the mass parameters $m_u,m_c,m_t$
and one of the parameters $m_d,m_s,m_b$ be negative.

Given arbitrary matrices $M_u$, $M_d$ or equivalently $D_u$, $D_d$,
$V$, one has to find a unitary matrix $W$ satisfying
(\ref{system1}). In general, it is not an easy task to find a closed
analytical solution for the system of Eqs.(\ref{system1}). In section
\ref{sec:numeric} we will give some numerical examples.  Next, we
shall study some special cases where a simple analytical solution for
$W$ can be found.

Let us consider first the special case of $V=\one$. In this case, $X=W$ and
the solution of the system of equations (\ref{system1}) is straightforward:
\begin{eqnarray}
|W_{11}|^2=\frac{m_c\,m_b-m_s\,m_t}{\Delta}, \nl
|W_{21}|^2=\frac{m_d\,m_t-m_u\,m_b}{\Delta}, \nl
|W_{31}|^2=\frac{m_u\,m_s-m_d\,m_c}{\Delta},\label{solcase1}
\end{eqnarray}
where
\begin{equation}
\Delta=(m_t-m_u)(m_b-m_s)-(m_t-m_c)(m_b-m_d). \label{Delta}
\end{equation}

Next we consider another example, where it is also possible to find an
analytical solution for W. We choose $V$ closer to a realistic CKM matrix:
\begin{equation}
V=\left(
\begin{array}{ccc}
 \cos \theta & \sin \theta & 0 \\
 -\sin \theta & \cos \theta & 0 \\
 0 & 0 & 1
\end{array}
\right)\, .
\end{equation}

In this case, Eqs.(\ref{X}) lead to:
\begin{eqnarray}
|X_{11}|^2 & = & \cos^2 \theta \: |W_{11}|^2 + \sin^2 \theta \:
 |W_{21}|^2 - \sin 2\theta \:W_{11}\:W_{21}, \nl \nl
|X_{21}|^2 & = & \sin^2 \theta \: |W_{11}|^2 + \cos^2 \theta \:
 |W_{21}|^2 + \sin 2\theta\: W_{11}\:W_{21}, \nl \nl
|X_{31}|^2 & = & |W_{31}|^2,
\end{eqnarray}
where we have assumed $W_{i1}\, (i=1,2,3)$ to be real.

Using unitarity, we can write
\begin{equation}
\begin{array}{l}
(m_u-m_t)|W_{11}|^2+(m_c-m_t)|W_{21}|^2+m_t=0, \\ \\ (m_d\:\cos^2
 \theta+m_s\:\sin^2\theta-m_b)|W_{11}|^2 \\ \hspace{1cm}+ (m_d\:\sin^2
 \theta+m_s\:\cos^2\theta-m_b)|W_{21}|^2\\ \hspace{2cm} + (m_s-m_d)\sin
 2\theta\:W_{11}W_{21} + m_b=0\, .
\label{system2}
\end{array}
\end{equation}

The solution of the above system can be parametrized as:
\begin{eqnarray}
\sqrt{m_t-m_u}\,W_{11}=\sqrt{m_t}\, \cos \varphi, \nl
\sqrt{m_t-m_c}\,W_{21}=\sqrt{m_t}\, \sin \varphi. \label{solcase2}
\end{eqnarray}
Denoting
\begin{eqnarray}
a&=&m_b-(m_b-m_d\sin^2\theta-m_s\cos^2\theta)\,\frac{m_t}{m_t-m_c}, \nl
b&=&(m_s-m_d)\,\frac{m_t \sin 2\theta}{\sqrt{(m_t-m_u)(m_t-m_c)}}, \nl
c&=&m_b-(m_b-m_d\cos^2\theta-m_s\sin^2\theta)\,\frac{m_t}{m_t-m_u},
\end{eqnarray}
and introducing $z\equiv \tan \varphi$, the solution is simply given by the
quadratic equation
\begin{equation}
a\,z^2+b\,z+c=0. \label{tanphi}
\end{equation}
In particular for $\theta=0$, i.e. $V=\one$, we recover the result of
Eqs.(\ref{solcase1}).

In order to emphasize that one can obtain the (1,1) zero
simultaneously in $M'_u$, $M'_d$, starting from arbitrary $M_u$,
$M_d$, we consider next the case of a completely unrealistic
$V_{CKM}$, namely:

\begin{equation}
V=\left(
\begin{array}{ccc}
1 & 0 & 0 \\ 0 & \cos \theta & \sin \theta  \\ 0 & \sin \theta & -\cos \theta
\end{array}
\right).
\end{equation}

It can be readily shown that the solution of the system
(\ref{system1}) can be obtained from the previous case by the
substitutions $ m_d\,\longleftrightarrow\,m_b, \quad
|W_{11}|^2\longleftrightarrow\,|W_{31}|^2, \quad
m_u\,\longleftrightarrow\,m_t$ in Eqs.(\ref{system2}) and the
corresponding changes in Eqs.(\ref{solcase2})-(\ref{tanphi}).

\subsection{The (1,3) WB zero}

We have seen that it is possible in general to perform a WB transformation on
the quark mass matrices such that the matrix elements in the position (1,1)
are equal to zero.  A natural question to ask is whether one can get
additional WB zeros, besides the ones already obtained for the (1,1) matrix
elements.  In order to show that this is indeed possible, let us assume that
one has already performed a WB transformation so that
$(M_u)_{11}=(M_d)_{11}=0$. It can be readily seen that there exists a second
WB transformation that keeps $(M'_u)_{11}=(M'_d)_{11}=0$ and leads to
$(M'_d)_{13}=(M'_d)_{31}=0$. Such a transformation is defined by
Eqs.(\ref{WBT1}), where
\begin{equation}
W=\left(
\begin{array}{ccc}
 1 & 0 & 0 \\ 0 & \cos \theta & -e^{i \varphi}\sin \theta \\ 0 & e^{-i
 \varphi}\sin \theta & \cos \theta \label{WBT2}
\end{array}
\right),
\end{equation}
with $\theta$ and $\varphi$ given by
\begin{equation}
\tan \theta = \left|\frac{(M_d)_{13}}{(M_d)_{12}}\right|,\quad \varphi
= \arg (M_d)_{13}-\arg (M_d)_{12}.
\end{equation}

Once this WB transformation is performed, we are left with the
following WB zero forms:
\begin{equation}
M'_u=\left(
\begin{array}{ccc}
  0 & \ast & \ast \\
  \ast & \ast & \ast \\
  \ast & \ast & \ast
\end{array}
\right),\:
M'_d=\left(
\begin{array}{ccc}
  0 & \ast & 0 \\
  \ast & \ast & \ast \\
  0 & \ast & \ast
\end{array}
\right). \label{texture3}
\end{equation}

At this point it is worth mentioning that the construction of mass
matrices with two off-diagonal zeros in one matrix and one
off-diagonal zero in the other matrix is straightforward, if one
starts in the basis where one of the matrices is diagonal (cf.
Eq.(\ref{mudiag})) and then perform the WB given in (\ref{WBT2}). All
the cases with two off-diagonal zeros in one sector and one
off-diagonal zero in the other can be studied following this
procedure. We are however concerned with the more interesting case
when the zeros are also present in the diagonal matrix elements.

Next we address the question whether it is possible to obtain an
additional zero at positions (1,3) and (3,1) of the up quark mass
matrix leading to the following structure
\begin{equation}
M''_u=\left(
\begin{array}{ccc}
  0 & \ast & 0 \\
  \ast & \ast & \ast \\
  0 & \ast & \ast
\end{array}
\right),\:
M''_d=\left(
\begin{array}{ccc}
  0 & \ast & 0 \\
  \ast & \ast & \ast \\
  0 & \ast & \ast
\end{array}
\right). \label{texture4}
\end{equation}

The structure of the mass matrices given by Eq.(\ref{texture4}) has the
interesting feature of exhibiting a parallel structure for the up and down
mass matrices and having ten independent parameters. This can be seen by
noting that one can make quark phase redefinitions which render $M''_u$ real,
while leaving $M''_d$ with two phases. Thus the number of independent
parameters in $M''_u$, $M''_d$ coincides with the number of physical
parameters contained in the up and down quark mass matrices. Therefore, it is
pertinent to ask whether one can always reach the form of Eq.(\ref{texture4}),
starting from arbitrary matrices $M_u$, $M_d$, through a WB transformation. It
can be easily shown that this is generally not the case.

Let us assume that one starts from the basis of Eq.(\ref{mudiag}) and
take the special case $V=\one$. Let us consider a WB transformation
performed by a unitary matrix $W$, as defined by Eq.(\ref{WBT1}). In
order for this WB transformation to lead to the structure of
Eq.(\ref{texture4}), $W$ will have to satisfy not only
Eqs.(\ref{solcase1}), but also the relations:
\begin{eqnarray}
(m_t-m_u)W^{\ast}_{11}W_{13}+(m_t-m_c)W^{\ast}_{21}W_{23}&=&0, \nl
(m_b-m_d)W^{\ast}_{11}W_{13}+(m_b-m_s)W^{\ast}_{21}W_{23}&=&0.
\end{eqnarray}
It is clear that there is no unitary matrix $W$ satisfying the above
relations for arbitrary quark masses. Therefore the form of
Eq.(\ref{texture4}) does not reflect, in general, a WB choice; on the
contrary, it implies constraints on quark masses and
mixings. Nevertheless, there are CKM matrices and quark masses
consistent with the present experimental data, for which it does exist
a WB transformation leading to four WB zeros as in
Eq.(\ref{texture4}).  In the next section we will present a numerical
example where the matrix $W$ is explicitly given.

\section{Numerical Examples}
\label{sec:numeric}
To present our numerical examples we shall use the standard
parametrization of the CKM matrix advocated in \cite{pdg},

\begin{equation}
V=\left(
\begin{array}{ccc}
 c_{12}c_{13} & s_{12}c_{13} & s_{13}\,e^{-i\delta_{13}} \\
 -s_{12}c_{23}-c_{12}s_{23}s_{13}\,e^{i\delta_{13}} &
  c_{12}c_{23}-s_{12}s_{23}s_{13}\,e^{i\delta_{13}} & s_{23}c_{13} \\
  s_{12}s_{23}-c_{12}c_{23}s_{13}\,e^{i\delta_{13}} &
 -c_{12}s_{23}-s_{12}c_{23}s_{13}\,e^{i\delta_{13}} & c_{23}c_{13} \label{vpdg}
\end{array}
\right),
\end{equation}
with $c_{ij}=\cos \theta_{ij}$ and $s_{ij}=\sin \theta_{ij}$, $i,j=1,2,3$.
Since $c_{13}$ is known experimentally to be very close to unity \cite{pdg},
then one has $V_{us} \simeq s_{12}, V_{ub} \simeq s_{13}, V_{cb} \simeq
s_{23}$.

We shall take the following quark mass values (in GeV) given at the
electroweak ($\mu=M_Z$) scale:
\begin{equation}
\begin{array}{lll}
|m_u|=0.0025,&|m_c|=0.6,&|m_t|=174,\\
|m_d|=0.004,&|m_s|=0.08,&|m_b|=3.
\end{array}
\label{qmasses}
\end{equation}

Let us consider first a CKM matrix consistent with the present
experimental data. Assuming $V_{us}=0.22,V_{ub}=0.0036,V_{cb}=0.04$
and $\delta_{13}=\pi/2$, we find from (\ref{vpdg}) that
\begin{equation}
|V| = \left(
\begin{array}{ccc}
 0.9755 & 0.22 & 0.0036 \\
 0.2198 & 0.9747 & 0.04 \\
 0.0095 & 0.039 & 0.9992
\end{array}
\right).
\end{equation}

Now it is easy to find the WB transformation (\ref{WBT1}) such that
the new quark mass matrices contain four WB zeros at positions (1,1)
and (1,3). Taking $m_u<0,m_d<0$ in (\ref{qmasses}), one has
\begin{equation}
W = \left(
\begin{array}{ccc}
 -0.9979 & -0.0096-0.043\,i & -0.0459+0.0029\,i \\
 0.0137-0.0628\,i & -0.6946 & -0.1076+0.7084\,i \\
 0.0002 & -0.1078-0.7099\,i & 0.696
\end{array}
\right),
\end{equation}
and the quark mass matrices are
\begin{equation}
M'_u = \left(
\begin{array}{ccc}
 0 & -0.0118-0.0543\,i & 0 \\
 -0.0118+0.0543\,i & 89.99 & -13.0092+85.6774\,i \\
 0 & -13.0092-85.6774\,i & 84.6076
\end{array}
\right),
\end{equation}
\begin{equation}
M'_d = \left(
\begin{array}{ccc}
 0 & 0.0242-0.0079\,i & 0 \\
 0.0242+0.0079\,i & 1.6007 & -0.3319+1.4223\,i \\
 0 & -0.3319-1.4223\,i & 1.4753
\end{array}
\right).
\end{equation}

In order to illustrate the fact that the form of Eq.(\ref{texture3})
only reflects a choice of WB, we give next an example where we take
the same realistic values for quark masses given by
Eq.(\ref{qmasses}), but choose a completely unrealistic CKM
matrix. Taking $\theta_{12}=\theta_{13}=\theta_{23}=\pi/6$ and
$\delta_{13}=\pi/2$ in (\ref{vpdg}) we obtain:
\begin{equation}
|V| = \left(
\begin{array}{ccc}
 3/4 & \sqrt{3}/4 & 1/2  \\ \\
 \sqrt{15}/8 & \sqrt{37}/8 & \sqrt{3}/4\\ \\
 \sqrt{13}/8 & \sqrt{15}/8  & 3/4
\end{array}
\right).
\end{equation}

It is easy to construct a unitary matrix $W$ such that the WB
transformation given in Eq.(\ref{WBT1}) transforms the quark mass
matrices into the WB zero form of Eq.(\ref{texture3}). Assuming for
this case $m_c<0,m_s<0$ in (\ref{qmasses}) we find:
\begin{equation}
W = \left(
\begin{array}{ccc}
 -0.7345 & 0.1049-0.4481\,i & -0.4986\,i \\
 0.6774\,i & 0.5272+0.1134\,i & 0.5002+0.009\,i \\
 -0.0397\,i & 0.7052-0.0066\,i & -0.6908+0.1544\,i
\end{array}
\right),
\end{equation}
\begin{equation}
M'_u = \left(
\begin{array}{ccc}
 0 & -0.0004+5.0843\,i & -1.0697-4.5653\,i \\
 -0.0004-5.0843\,i & 86.373 & -85.1037+18.1793\,i \\
 -1.0697+4.5653\,i & -85.1037-18.1793\,i & 87.0295
\end{array}
\right),
\end{equation}
\begin{equation}
M'_d = \left(
\begin{array}{ccc}
 0 & -0.0309+0.314\,i & 0 \\
 -0.0309-0.314\,i & 2.9144 & -0.1218+0.3768\,i \\
 0 & -0.1218-0.3768\,i & 0.0096
\end{array}
\right).
\end{equation}

\section{Four texture zeros and the CKM matrix}
\label{sec:ckm}
The four texture zero form of Eq.(\ref{texture4}) is specially
interesting because the latest low energy data \cite{pdg} seems to
rule out all six and five texture zero quark mass matrices. Let us
assume the following up and down quark mass matrices:
\begin{equation}
M_u = \left(
\begin{array}{ccc}
 0 & a_u & 0 \\
 a_u & b_u & c_u \\
 0 & c_u & d_u
\end{array}
\right), \: M_d = P\,\left(
\begin{array}{ccc}
 0 & a_d & 0 \\
 a_d & b_d & c_d \\
 0 & c_d & d_d
\end{array}
\right)\,P^\dagger\, ,\label{texture-4}
\end{equation}
where $P=\diag\,(e^{i \alpha_1},e^{i \alpha_2},1)$. The matrix
elements of a real matrix of the type

\begin{equation}
M = \left(
\begin{array}{ccc}
 0 & a & 0 \\
 a & b & c \\
 0 & c & d
\end{array}
\right),
\end{equation}
are related to the eigenvalues $m_i,\, i=1,2,3$, through the invariants:
 \begin{eqnarray}
 \trace (M) &=& b+d=m_1+m_2+m_3\, , \nl
 \det (M) &=& -a^2 d = m_1 m_2 m_3\, , \nl
 \chi (M) &=& b d - a^2 - c^2 = m_1 m_2 +m_1 m_3 +m_2 m_3\, .
 \end{eqnarray}

Without loss of generality we can order the eigenvalues $m_i$ in such
a way that $|m_1| < |m_2| < |m_3|$ and also assume that $m_1<0$,
$m_3>d>m_2>0$. In this case the unitary matrix $U$ which diagonalizes
$M$ is given by
\begin{equation}
U = \left(
\begin{array}{ccc}
 \sqrt{\frac{m_2 m_3 (d-m_1)}{d\, (m_2-m_1) (m_3-m_1)}} &
  \sqrt{\frac{m_1 m_3 (m_2-d)}{d\, (m_2-m_1) (m_3-m_2)}} &
    \sqrt{\frac{m_1 m_2 (d-m_3)}{d\, (m_3-m_1) (m_3-m_2)}} \\
  \\
  -\sqrt{\frac{m_1 (m_1-d)}{(m_2-m_1) (m_3-m_1)}} &
   \sqrt{\frac{(d-m_2) m_2}{(m_2-m_1) (m_3-m_2)}} &
   \sqrt{\frac{m_3 (m_3-d)}{(m_3-m_1) (m_3-m_2)}}\\
  \\
  \sqrt{\frac{m_1 (d-m_2) (d-m_3)}{d\, (m_2-m_1) (m_3-m_1)}} &
   -\sqrt{\frac{m_2 (d-m_1) (m_3-d)}{d\, (m_2-m_1) (m_3-m_2)}}  &
    \sqrt{\frac{m_3 (d-m_1) (d-m_2)}{d\, (m_3-m_1) (m_3-m_2)}}
\end{array}
\right). \label{matrixU}
\end{equation}

If we assume the mass hierarchy $|m_1| \ll m_2 \ll m_3$, then the
matrix $U$ can be simplified:
\begin{equation}
U \simeq \left(
\begin{array}{ccc}
 1 &
  \sqrt{\frac{m_1 (m_2-d)}{d\, m_2}}&
    \sqrt{\frac{m_1 m_2 (d-m_3)}{d\, m^2_3}} \\
 \\
  -\sqrt{\frac{d |m_1|}{m_2 m_3}} &
   \sqrt{\frac{d-m_2 }{m_3}} &
    \sqrt{\frac{m_3-d}{m_3}} \\
  \\
  \sqrt{\frac{m_1 (d-m_2) (d-m_3)}{d\, m_2 m_3}} &
   -\sqrt{\frac{m_3-d}{m_3}}  &
    \sqrt{\frac{d-m_2}{m_3}}
\end{array}
\right), \label{matrixU1}
\end{equation}

If we make the additional assumption that $d \approx m_3$ then the
matrix $U$ is even further simplified:
\begin{equation}
U \approx \left(
\begin{array}{ccc}
 1 &
  \sqrt{\frac{|m_1|}{m_2}}&
    \sqrt{\frac{m_1 m_2 (d-m_3)}{m^3_3}} \\
\\
  -\sqrt{\frac{|m_1|}{m_2}} &
   1 &
    \sqrt{\frac{m_3-d}{m_3}}\\
\\
  \sqrt{\frac{m_1 (d-m_3)}{m_2 m_3}} &
   -\sqrt{\frac{m_3-d}{m_3}}  &
    1
\end{array}
\right), \label{matrixU2}
\end{equation}

Now one can easily find the CKM matrix $V=U^\dagger_u\,P\,U_d$. In
particular, using the matrix form (\ref{matrixU2}) for $U_u,U_d$ we
obtain the successful prediction \cite{weinberg},

\begin{equation}
 |V_{us}| \simeq \left| \sqrt{\frac{|m_d|}{m_s}}-
 \sqrt{\frac{|m_u|}{m_c}}\,e^{i (\alpha_2-\alpha_1)} \right|\, .
\label{vus}
 \end{equation} 
Moreover the following ratios are obtained:
\begin{equation}
\frac{|V_{ub}|}{|V_{cb}|} \simeq \sqrt{\frac{|m_u|}{m_c}}\, , \quad
\frac{|V_{td}|}{|V_{ts}|} \simeq \sqrt{\frac{|m_d|}{m_s}}\, , \label{ratios}
\end{equation}
which are common to some models with nearest-neighbour mixing
\cite{albright,branco}.

\section{Discussion and Conclusions}

In this letter we address the question of the physical meaning of
texture zeros in the up and down quark mass matrices. We have pointed
out that some of the zeros included in most of the \emph{Ans\"{a}tze}
proposed in the literature only reflect a choice of WB, in the sense
that one can obtain them starting from arbitrary mass matrices $M_u$,
$M_d$, by making an appropriate WB transformation. This is the case of
the three WB zero structure of Eq.(\ref{texture3}). We have then
examined the four texture zero structure of Eq.(\ref{texture4}), which
is specially important in view of the fact that the latest low energy
data disfavours \cite{ali,randhawa} all six and five texture zero
quark mass matrices. We have also pointed out that not all matrices
$M_u$, $M_d$ can be put in the form of the Eq.(\ref{texture4}),
through a choice of WB. Nevertheless, we have shown there are cases of
quark masses and mixings for which such a WB transformation indeed
exists. This four texture zero structure has ten independent
parameters and thus one would expect that it would have a rather limit
predictive power, since there are also ten physical parameters
corresponding to the six quark masses and four CKM
parameters. However, as we have seen in section \ref{sec:ckm}, the
four texture zero \emph{Ansatz} of Eq.(\ref{texture4}), together with
some assumptions which include the quark mass hierarchies, does lead
to successful predictions for $V_{CKM}$ such as those of
Eqs.(\ref{vus}),(\ref{ratios}). Finally, it should be mentioned that
another attractive feature of the four texture zero \emph{Ansatz}
considered here is the fact that it can successfully describe not only
the quark but also the lepton sector, in particular the charged lepton
and neutrino masses \cite{nishiura}. If the Dirac neutrino and
Majorana mass matrices, $M_D,M_R$, are assumed to have a zero texture
structure similar to the quark mass matrices, i.e. with zeros in
positions (1,1),(1,3) and (3,1), then the neutrino mass matrix
obtained via the see-saw mechanism, $M_\nu=-M_D^T\,M_R^{-1}\,M_D$,
will have the same structure since the latter transformation does not
affect the above zero textures.

Throughout this letter, we have used the fact that in the SM all WB are
equivalent. In theories beyond the SM, where family symmetries may be present,
not all WB are equivalent, since the symmetry may single out a particular WB.
Note however that even if one finds a family symmetry which leads
automatically to the four-texture zero structure of Eq.(\ref{texture4}), in
order to gain predictive power one has to find an additional mechanism to
explain the mass hierarchies. An especially attractive mechanism is the one
suggested by Froggatt and Nielsen \cite{froggatt} through a broken continuous
Abelian symmetry beyond the SM.

\acknowledgments One of us (D.E.C.) would like to thank the {\em Funda\c{c}\~{a}o de
Ci\^{e}ncia e Tecnologia} ({\em Segundo Quadro Comunit\'ario de Apoio}) for
financial support under the contract No. Praxis XXI/BD/9487/96.

\end{document}